\documentclass[prl,twocolumn,aps,showpacs,superscriptaddress]{revtex4}
\usepackage{graphics}
\usepackage{epsfig}
\usepackage{amsmath}
\usepackage{amssymb}
\usepackage{bm}
\usepackage{color}

\begin{document}

\title{Floppiness, cutting, and freezing: Dynamic critical scaling near isostaticity}
\author{Brian P. Tighe}
\affiliation{Delft University of Technology, Process \& Energy Laboratory, Leeghwaterstraat 44, 2628 CA Delft, The Netherlands}

\date{\today}

\begin{abstract}
The isostatic state plays a central role in organizing the response of many amorphous materials. We demonstrate the existence of a dynamic critical length scale in nearly isostatic spring networks that is valid both above and below isostaticity and at finite frequencies, and use scaling arguments to relate  the length scale to viscoelastic response. We predict theoretically and verify numerically how proximity to isostaticity controls the viscosity, shear modulus, and creep of  random networks.
\end{abstract}

\maketitle

Random networks of springs display two distinct phases distinguished by the presence or absence of {\em floppy modes} -- zero frequency motions that neither compress nor stretch springs. Floppiness is related to network structure via the mean coordination $z$ \cite{alexander}; networks with floppy modes reside below the {\em isostatic} coordination $z_c$. Many amorphous materials possess an isostatic state, including fiber networks, covalent glasses, foams, and emulsions. Hence the viscoelasticity of damped random networks, which remains poorly understood, could potentially provide insight into a broad class of materials, including how structure relates to response \cite{alexander,thorpe00,heussinger06,wyart08,ellenbroek09b,hatano09,vanhecke10,tighe11,broedersz10,broedersz11,wyart10,sheinman12}.

The dramatic impact of floppiness on response is apparent in numerical measurements of the creep compliance $J(s) = \gamma(s)/\sigma_0$, shown in Fig.~1; $\gamma(s)$ is the Laplace-transformed shear strain accrued after a small step stress $\sigma_0$.  At long times or vanishing $s$, hyperstatic networks ($z > z_c$) approach constant strain, $J \sim s^0$, while hypostatic networks ($z < z_c$),  approach constant strain rate, $J \sim 1/s$. Thus  networks with floppy modes are fluids, and those without are solids. Moreover, networks' elasticity and viscosity clearly vary with proximity to $z_c$. 

Together with prior work \cite{ellenbroek09b,wyart08,wyart10,tighe11,sheinman12}, data such as in Fig.~1 strongly suggest that the isostatic state is a nonequilibrium critical point. We will show this is indeed the case, and that proximity to $z_c$ controls viscoelasticity in random spring networks.  To fully understand dynamics near a critical point, it is essential to identify the associated diverging length scale \cite{hohenberghalperin}. While hyperstatic networks possess an ``isostatic length'' $\ell^* \sim 1/\Delta z$, where $\Delta z \equiv z - z_c$ \cite{tkachenko99,wyart05,silbert05,ellenbroek06,mailman11}, its derivation is only valid for quasistatic response above $z_c$. 

To explain data such as in Fig.~1, we must identify a length scale $\xi_\pm(\Delta z, \omega)$ valid not only above $z_c$, but also below and at finite frequency $\omega$ (or rate $s$). 
Here we determine $\xi_\pm$ and show that it is the size of a finite system in which elastic storage and viscous loss balance. We also provide scaling arguments showing that
 response functions such as $J(s)$ and the complex shear modulus $G^*(\omega)$ are controlled by $\xi_\pm$. Moreover, their rate and frequency dependence matches those found in jammed solids \cite{tighe11}, biopolymer networks \cite{broedersz10}, and athermal suspensions \cite{hatano09}.

\begin{figure}[tb]
\centering 
\includegraphics[clip,width=0.9\linewidth]{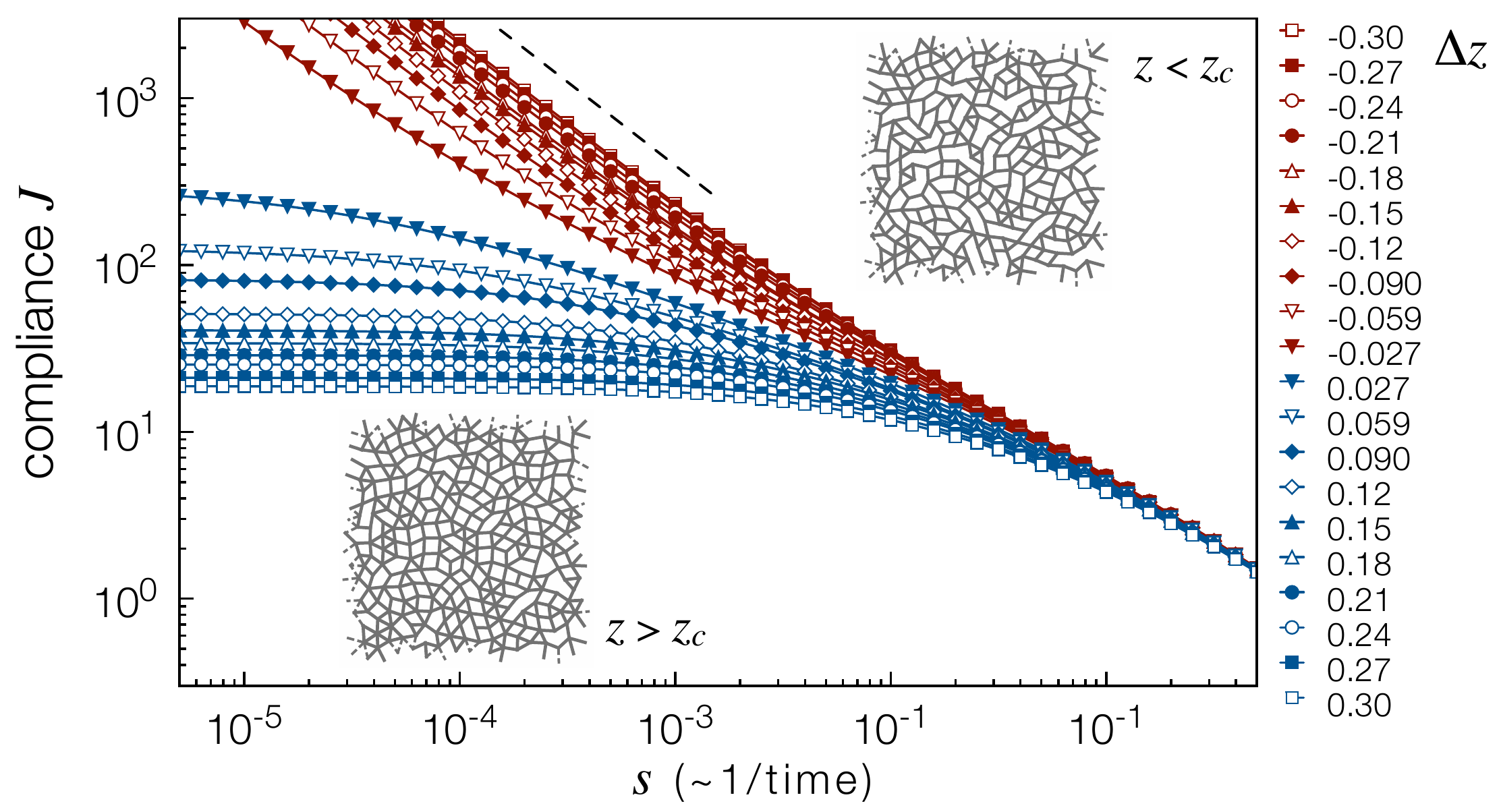}
\caption{Numerical measurement of creep compliance $J(s)$ in damped random spring networks for varying distance to isostaticity $\Delta z \equiv z - z_c$. For long time (low $s$), networks below isostaticity begin to flow, $J \sim 1/s$ (dashed line), while networks above isostaticity achieve a finite strain.}
\end{figure}

{\em Random spring networks.---} 
The critical coordination of central force networks with dimensionality $d$ is $z_c = 2d$ \cite{alexander}. We generate networks near $z_c$ according to the protocol of Ref.~\cite{ellenbroek09b}, which begins from an initial network with high coordination and randomly removes springs to reach a targeted $z$. The initial disordered network is derived from a numerically generated sphere packing. During dilution, springs are removed only if the affected nodes retain at least $d+1$ springs; we present numerical results for $d=2$, though the scalings we predict are independent of $d$. 
This rule guarantees that  all springs generically bear load in hyperstatic networks, and that even hypostatic networks are above connectivity percolation.

A network's nodes are connected by springs at their rest length with stiffness $k$. When nodes undergo displacements $\lbrace \vec U_i \rbrace$,  the force from node $j$ on node $i$, $\vec f_{ij} =  k  u^\parallel_{ij} \, \hat n_{ij}$, is along the unit vector $\hat n_{ij}$ pointing from $i$ to $j$ and proportional to their relative normal motion $u^\parallel_{ij} = (\vec U_j - \vec U_i)\cdot \hat n$. 
To introduce damping, we take the network to be immersed in an affinely flowing Newtonian fluid: under a pure shear $\hat \gamma(t)$, the fluid element at position $\vec x$  has displaced by $ \hat{ \gamma}(t) \vec x$ at time $t$. 
A node's velocity relative to the fluid, or non-affine velocity $ \dot{\vec U}^{\rm na}_i$, is damped by a viscous force $\vec F _i= - b  \dot{\vec U}^{\rm na}_i$, with damping coefficient $b$.
We set the stiffness $k$, the damping $b$, and the average spring length all to unity. Equations of motion and numerical details are given in the Supplementary Material.

{\em Gradients in the solid and fluid phases.---} 
Our scaling arguments are built on two simple relationships between relative normal motions $u^\parallel$ and non-affine motions $U^{\rm na}$. We first develop these relationships before turning to $\xi_\pm$.

In a solid, zero frequency shear stretches and compresses springs. The normal motion $u^\parallel_{ij}$ in spring $(ij)$is a discretization of $\vec \nabla_\parallel \cdot \vec U$, the local gradient of the displacement field $\vec U$ along $\hat n$.
We assume the gradient is dominated by $\vec U^{\rm na}$. 
For scaling purposes, we introduce a quantity $\lambda_s$, as yet unspecified, that relates the typical value of $u^\parallel$ and the typical non-affine motion $U^{\rm na}$:
\begin{equation}
 u^\parallel \sim \frac{ U^{\rm na} }{\lambda_s } \,.
\label{eqn:strain}
\end{equation}
Because $ u^\parallel$ and $U^{\rm na}$ are related by a gradient, $\lambda_s$ is a length scale characterizing the solid (non-floppy) phase.

A fluid possesses floppy modes and can perform zero frequency deformations without stretching springs. As there can be non-affine motion while $ u^\parallel $ is everywhere zero,  Eq.~(\ref{eqn:strain}) must break down.
Elastic forces in a fluid are instead induced by flow. For oscillatory driving at finite frequency $\omega$ (or relaxation at rate $s$), nodes have finite velocity and experience viscous forces $\lbrace \vec F_i \rbrace$. Since inertia is negligible for small $\omega$, these viscous forces must be balanced by elastic forces $\lbrace \vec f_{ij} \rbrace$ in the springs. 

The force balance equations  $\sum_j \vec f_{ij} = \vec F_i$ are a discrete counterpart of the continuum relation ${\rm div} \, \hat \sigma = \vec {\cal F}$, where $\vec {\cal F}$ is the viscous force density and the stress tensor 
$\hat \sigma$
is linear in the elastic forces. Thus viscous forces are related to gradients in the elastic forces,
and there is an undetermined length scale $\lambda_f$ relating the typical elastic force $f$ and viscous force $F$ in the fluid (floppy) phase,
\begin{equation}
F \sim \frac{f}{\lambda_f} \,,
\label{eqn:fb}
\end{equation}
or equivalently $u^\parallel \sim \omega \lambda_f U^{\rm na}$.

In the following, an important consideration will be the complex shear modulus $G^*(\omega) = 1/J(s)|_{s = \imath \omega} \equiv G'(\omega) + \imath G''(\omega)$. Its real and imaginary parts $G'$ and $G''$, known as the storage and loss moduli, are energy densities  associated with elastic storage and viscous dissipation, respectively. Their scaling  can be written
\begin{equation}
G' \sim \vec f \cdot \vec u \sim \left \lbrace
  \begin{array}{cl}
    \left(  \lambda_f U^{\rm na} \right)^2 \omega^2 & \text{with floppy modes} \\
    \left( U^{\rm na} / \lambda_s \right)^2 & \text{without floppy modes,} 
  \end{array} \right.
\label{eqn:storage}  
\end{equation}
where we have made use of Eqs.~(\ref{eqn:strain}) and (\ref{eqn:fb}), and
\begin{equation}
G'' \sim \vec F \cdot \vec U \sim (U^{\rm na})^2 \omega \,,
\label{eqn:loss}
\end{equation}
both with and without floppy modes. Note that because $G^*(\omega)$ and $J(s)$ are related, for scaling purposes the inverse time scales $\omega$ and $s$ are interchangeable.

{\em Generalized isostatic length scale.---}
\begin{figure}[tb]
\centering 
\includegraphics[clip,width=0.95\linewidth]{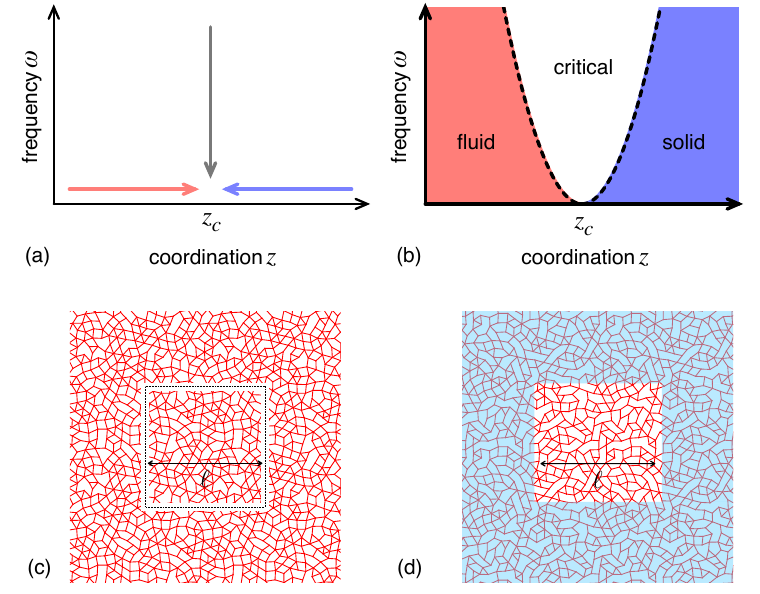}
\caption{(a) Three approaches to the critical point: along the $z$-axis from above and below, and along the frequency axis.
(b) The diverging length scale $\xi_\pm$ has three regimes, distinguished by comparing the frequency $\omega$ to the diverging time scale $\tau^* \sim 1/\Delta z^2$ (dashed curve).
(c) ``Cutting out'' a subsystem of size $\ell$. Springs crossing the dashed line are removed. (d) Creating a finite system by ``freezing.'' Nodes in the shaded region are not allowed to move.}
\end{figure}
We now introduce a scaling argument to identify a dynamic critical length scale $\xi_\pm(\Delta z,\omega)$. The subscript indicates the sign of $\Delta z$. 
In the conventional scenario for dynamic critical scaling near a critical point \cite{hohenberghalperin}, a diverging length scale displays three qualitatively distinct regimes  (Fig.~2a,b). Two regimes persist at zero frequency and can be probed by approaching $z_c$ from above and below at $\omega = 0$. 
The third or critical regime is inherently dynamical and can be probed by approaching $z_c$ along the frequency axis. 

Our approach is a generalization of the ``cutting argument,'' introduced in Refs.~\cite{tkachenko99} and \cite{wyart05}, which identifies a diverging length scale $\ell^* \sim 1/\Delta z$ associated with proximity to isostaticity. The derivation of $\ell^*$ is appropriate only for hyperstatic systems at zero frequency. 
Because we generalize the cutting argument, our length scale recovers $\ell^*$ in the quasistatic limit.

We  begin in the regime where $\ell^*$ obtains by approaching $z_c$ from above at $\omega = 0$. 
On long enough wavelengths it is reasonable to approximate a material as a continuum; $\ell^*$ is the length scale beyond which this approximation is valid. One way to identify $\ell^*$ is to consider the properties of a finite portion of the infinite system. A sample with linear size $\ell$ that has the same properties 
as the infinite system is larger than $\ell^*$. Alternatively, if the finite sample is qualitatively different -- for example, if it no longer resists quasistatic shear -- then $\ell < \ell^*$. Thus we must ask when such a difference first appears.

So motivated, we imagine ``cutting out'' a sample of an infinite random spring network with excess coordination $\Delta z$. To do this, we overlay a box of size $\ell$ on the infinite system and cut those springs that cross its boundary; see Fig.~2c. Cutting springs is destabilizing: cut enough and floppy modes will appear. When this happens the finite system must respond qualitatively differently to driving, as it cannot resist zero frequency shear. Therefore $\ell^*$ is the size of the box for which the first floppy mode appears. This occurs when the number of cut springs, proportional to the surface area $\ell^{d-1}$, exceeds the number of excess contacts, 
which is proportional to $\Delta z \, \ell^d$. Hence 
$(\ell^*)^{d-1} \sim \Delta z (\ell^*)^d$, or  $\xi_+(\Delta z,0) \equiv \ell^* \sim1/\Delta z$.

We now approach $z_c$ from below. Clearly repeating the cutting approach will not work in a system that already has floppy modes. 
As cutting is in essence the introduction of a free boundary, we now, instead, introduce a rigid boundary by ``freezing'' the position of nodes exterior to a box of linear size $\ell$ (see Fig.~2d). 
Freezing introduces stability: springs constrain floppy motion, and while each spring in the bulk is shared by two nodes, those springs connected at one end to a frozen node are not shared. Hence there are more constraints per degree of freedom in the finite system, and by choosing $\ell$ small enough we can remove all of its floppy modes. The counting is entirely analogous to the surface area-to-volume ratio described above and gives
$\xi_-(\Delta z,0) \sim1/(-\Delta z) $.

Lastly, we approach $z_c$ along the frequency axis. Recall that for the quasistatic 
systems, $\xi_\pm(\Delta z, 0)$ is the length scale for which a rigid system becomes floppy, or vice versa, under the influence of a boundary. As deformations of random spring networks at finite frequency both store and dissipate energy, to proceed in analogy to the quasistatic case we should ask for what length scale storage and dissipation are comparable. A length scale can be imposed either by cutting or freezing.

Let us first create a finite isostatic system by freezing.  The frozen system has no floppy modes, so low frequency storage will dominate loss. We assume that the system organizes its response so as to minimize the stored energy $G' \sim (U^{\rm na}/\lambda_s)^2$. This is achieved when $\lambda_s$ is as long as possible, $\lambda_s \sim \ell$ \cite{footnote}. By Eqs.~(\ref{eqn:storage}) and (\ref{eqn:loss}) the ratio of loss to storage is then $G''/G' \sim 1/\ell^2 \omega $, which is of order unity when
$\ell = \xi_-(0,\omega) \sim  1/\omega^{1/2}$. We identify this length scale with $\xi_-$ because it was reached via  freezing.
Alternatively, we can create a finite system by cutting an isostatic network. As cutting introduces floppy modes, low frequency loss will dominate storage. We assume that the response now minimizes the dissipated energy, $G'' \sim (U^{\rm na})^2\omega \sim  f^2/\lambda_f^2 \,\omega$, which is achieved if $\lambda_f$ is, likewise, identified with $\ell$ \cite{footnote}. The ratio $G''/G' \sim \ell^2 \omega$ is again of order unity when $\ell = \xi_+(0,\omega) \sim 1/\omega^{1/2} $.
Interestingly, displacement correlations in normal modes of undamped hyperstatic networks also decay as $1/\omega^{1/2}$ \cite{wyart10}.

In summary, we have used cutting and freezing arguments to infer a dynamic critical length scale $\xi_\pm(\Delta z, \omega)$. The length scale has two branches, a solid branch for $\Delta z > 0$ and a fluid branch for  $\Delta z < 0$, which together can be expressed succinctly as
\begin{equation}
\xi_\pm (\Delta z, \omega) \sim \left \lbrace
\begin{array}{cl}
\pm 1/\Delta z & \omega \tau^* \ll 1 \\
1/\omega^{1/2} & \omega \tau^* \gg 1 \,.
\end{array} \right.
\label{eqn:xipm}
\end{equation}
The crossover is controlled by the diverging time scale $\tau^* \sim 1/\Delta z^2$, which follows from balancing the scaling of $\xi_\pm$ in different regimes.
Because both branches of $\xi_\pm$ scale as $1/\omega^{1/2}$ near $z_c$, we can speak meaningfully of a single dynamic critical regime; see Fig.~2b. $\xi_\pm$ is frequency-independent outside the critical regime.

{\em Predicting response.---}
We now consider infinite or periodic systems.  Given $\xi_\pm$, we can construct a scaling argument for the response of nearly isostatic networks. 

The key step is to identify the gradient scales $\lambda_s$ and $\lambda_f$ in Eqs.~(\ref{eqn:strain}) and (\ref{eqn:fb}) with $\xi_+$ and $\xi_-$, respectively. 
This ansatz, which is consistent with our approach to finite systems, can be directly tested in numerics. If $\xi_\pm$ scales as in Eq.~(\ref{eqn:xipm}), then data from each branch will collapse when plotting $\Delta z^a \, (u^\parallel/U^{\rm na})$ and $\Delta z^a (f/F)$ versus $\omega/\Delta z^b$ for $a = 1$ and $b = 2$. As shown in Fig.~3a, there is excellent data collapse for $a = 1.1(1)$ and $b = 2.1(1)$, in good agreement with predictions. This confirms both the identification of $\lambda_f$ and $\lambda_s$ with $\xi_\pm$ and the scaling arguments that led to $\xi_\pm$.  

\begin{figure}[tb]
\centering 
\includegraphics[clip,width=1.0\linewidth]{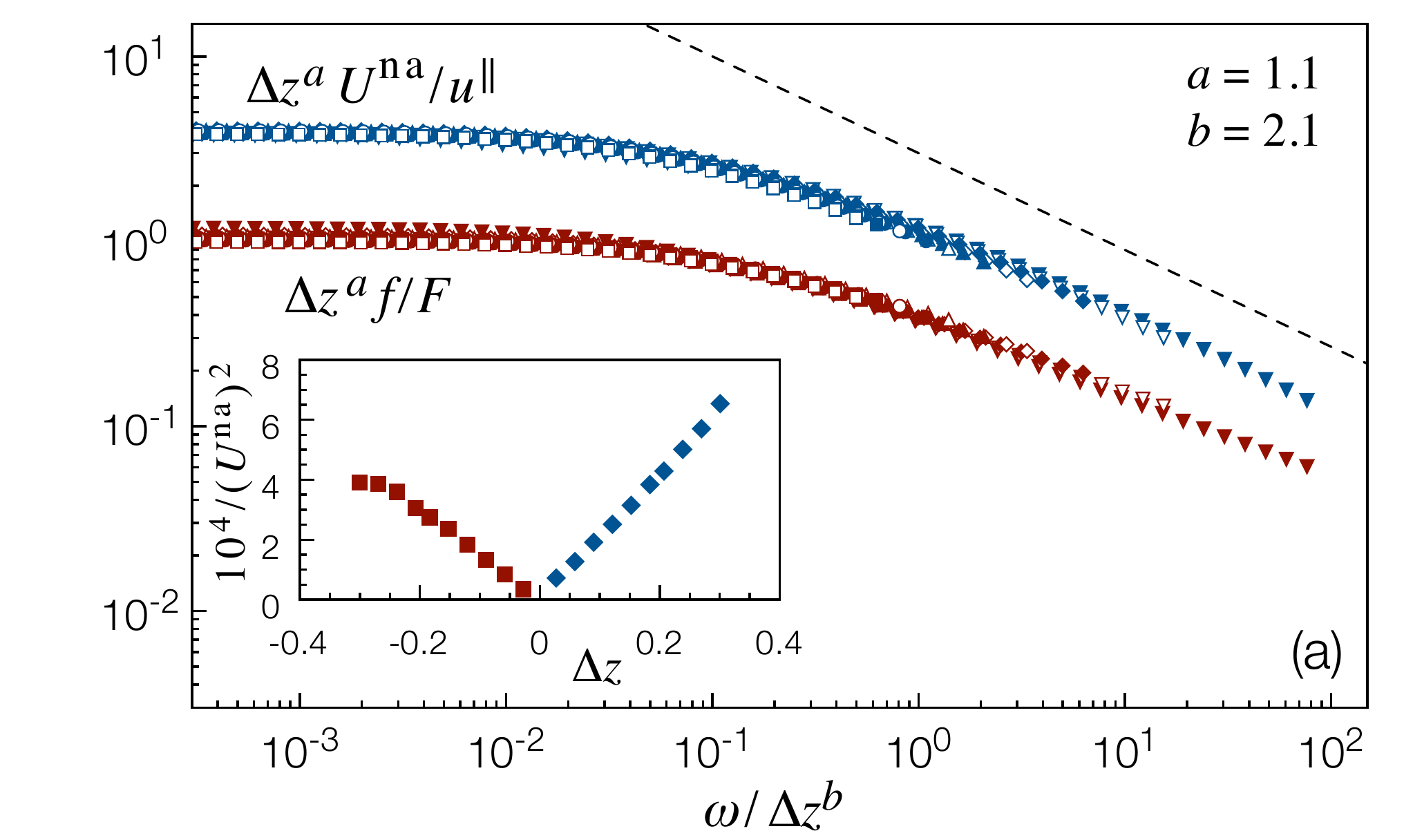}
\includegraphics[clip,width=1.0\linewidth]{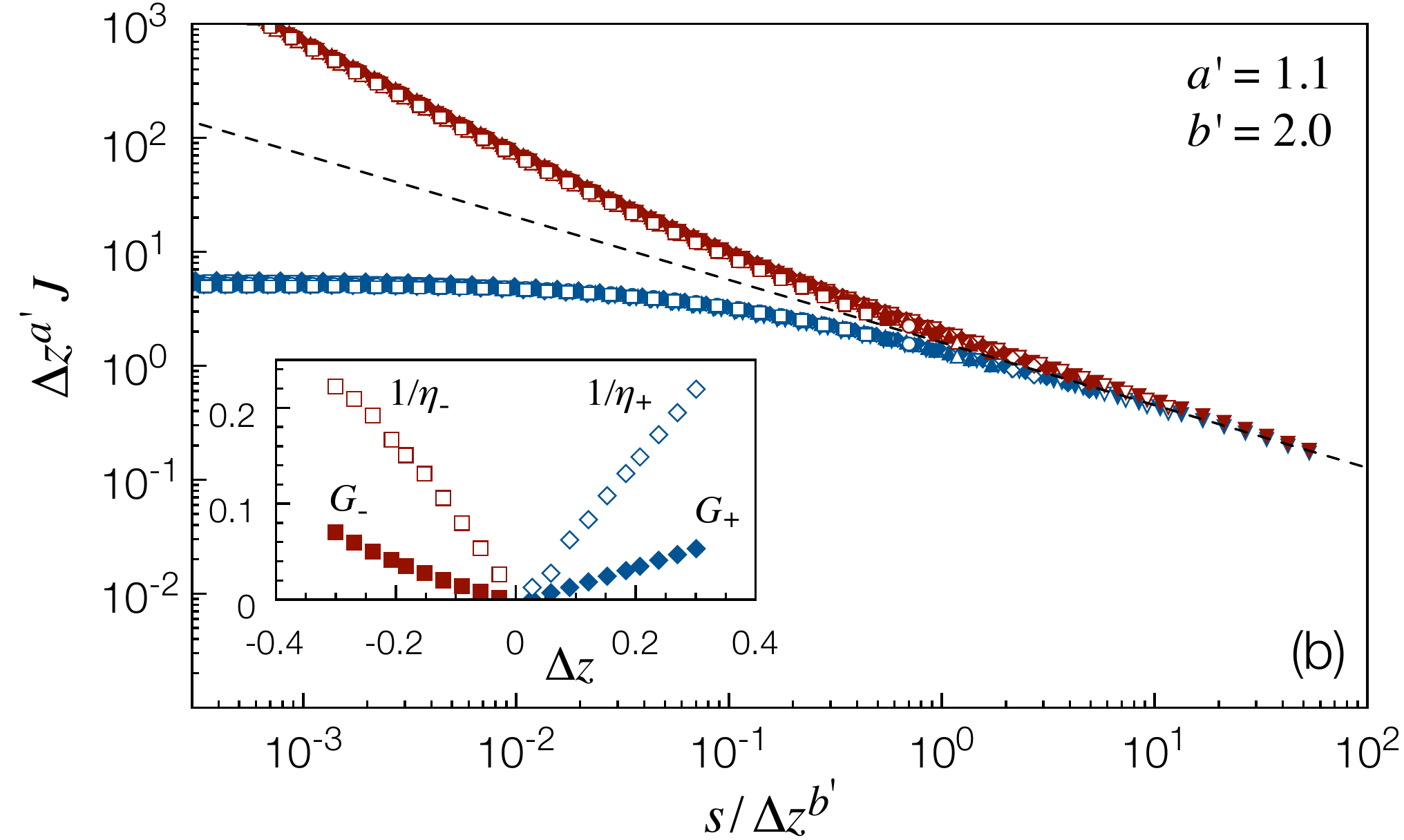}
\caption{Scaling collapse of data from simulated networks. Legend as in Fig.~1. 
(a) The diverging length scale $\xi_\pm$, determined from ratios of mean non-affine motion to mean relative normal motion and  mean elastic force to mean viscous force as a function of driving frequency $\omega$. The dashed curve has slope $a/b$. Inset: $1/(U^{\rm na})^2$ vanishes linearly at $z_c$ in the zero frequency limit. 
(b) Scaling collapse of the compliance data in Fig.~1. The dashed curve has slope $a'/b'$. Inset: Elastic and viscous coefficients in the zero frequency limit.
}
\end{figure}

Having identified $\lambda_s$ and $\lambda_f$, Eqs.~(\ref{eqn:storage}) and (\ref{eqn:loss}) relate the complex shear modulus to the typical magnitude of non-affine motion $U^{\rm na}$, which remains to be determined.
As non-affinity is a form of fluctuation, it is reasonable to expect its divergence on approach to the critical point \cite{hohenberghalperin}, $U^{\rm na} \sim \xi_{\pm}^\nu $. The exponent $\nu$ can be determined from the dynamic viscosity in hypostatic networks -- its scaling $\eta_0 \sim 1/(-\Delta z)$ follows from a straightforward calculation (see Supplementary Material). Physically, the viscosity diverges because low frequency motions of hypostatic networks strongly project on their floppy modes; $\eta_0^{-1}$ is controlled by the floppy mode number density, which is proportional to $z_c - z$.
Recalling that $\eta_0 = \lim_{\omega \rightarrow 0} G''/\omega$ and comparing to Eq.~(\ref{eqn:loss}), we conclude that $\nu = 1/2$; this prediction is confirmed in the inset to Fig.~3a.

With an expression for $\xi_\pm$ and the value of $\nu$, we have now determined the scaling of $G'$ and $G''$. These can be expressed in an elegant and compact form by introducing elastic and viscous coefficients
$G_\pm \sim 1/\xi_\pm$ and $\eta_\pm \sim \xi_\pm$.
For hyperstatic networks Eqs.~(\ref{eqn:storage}) and (\ref{eqn:loss}) become
\begin{equation}
G' \sim G_+ \,\,\,\,\,\, {\rm and} \,\,\,\,\,\, G'' \sim \eta_+ \omega \,.
\end{equation}
These are the moduli of the simplest possible viscoelastic solid,  the Kelvin-Voigt solid -- elastic and viscous elements with coefficients $G_+$ and $\eta_+$ connected in parallel. Similarly, for hypostatic systems we recover the low frequency moduli of elements $G_-$ and $\eta_-$ connected in series, i.e.~a Maxwell fluid,
\begin{equation}
G' \sim (\eta_- \omega)^2/G_-  \,\,\,\,\,\, {\rm and} \,\,\,\,\,\, G'' \sim \eta_- \omega \,.
\end{equation}
Thus a network's complex shear modulus is always described by an elastic element $G_\pm$ and a viscous dashpot $\eta_\pm$. Above isostaticity they are wired as a solid, below isostaticity they are wired as a fluid.

{\em Testing predictions.---} In addition to the direct test of Fig.~3a, we provide two further tests of our predictions. The first probes the zero frequency limit and confirms the scaling of $G_\pm$, which is given by $\lim_{\omega \rightarrow 0} G'$ above $z_c$ and $\lim_{\omega \rightarrow 0} (G'')^2/G'$ below, and $\eta_\pm = \lim_{\omega \rightarrow 0} G''/\omega$, which is valid on both sides of $z_c$.
Fig.~3b (inset) shows that, as predicted, both $G_\pm$ and $1/\eta_\pm$ grow linearly with $\Delta z$ on both sides of the transition.
 
Our second test revisits the compliance data of Fig.~1; by collapsing $J(s)$, we verify the scaling of $\xi_\pm$ on finite time scales. In hyperstatic networks the compliance is dominated by storage, $J \sim 1/G_+ \sim \xi_+(\Delta z,s)$, while in hypostatic systems it is controlled by loss, $J \sim 1/\eta_-s  \sim 1/[\xi_-(\Delta z,s)\, s]$. Therefore plotting $\Delta z^{a'} \, J$ versus $s/\Delta z^{b'}$ for $a'=1$ and $b' = 2$ should collapse data on both sides of the critical point to a two-branched master curve with a critical regime $J \sim 1/s^{1/2}$. We find excellent collapse for $a' = 1.1(1)$ and $b' = 2.0(1)$ (Fig.~3b), confirming these predictions. The $1/s^{1/2}$ divergence implies that strain grows anomalously with the square root of time before achieving constant strain or strain rate. This is reminiscent of Andrade creep in metals, although the Andrade creep exponent is $1/3$ rather than $1/2$ \cite{miguel02}.

{\em Conclusions.---} 
We have predicted, tested, and confirmed the form of a diverging dynamic length scale near isostaticity. When combined with simple scaling arguments, this length scale correctly predicts the viscoelastic response of nearly isostatic networks.
At frequencies low compared to $1/\tau^*$, $\xi_\pm$ is controlled by the distance to isostaticity, networks behave as simple viscoelastic solids or fluids, and their elastic and viscous moduli are set by $\Delta z$. At higher frequencies there is a critical regime in which $\xi_\pm$ is determined by the time scale with which the system is driven, storage and loss are comparable, and the distinction between solid and fluid is blurred. 

Though we have treated spring networks, our predictions, in particular the rate-dependence of the compliance and shear modulus, match prior results for $G^*$ in jammed solids \cite{tighe11} and biopolymer networks \cite{broedersz10} and the relaxation modulus in athermal suspensions \cite{hatano09}. This suggests the broader applicability of the scaling arguments presented here, which we expect can be extended to incorporate additional physics, including bending stiffness \cite{heussinger06,broedersz11}, finite strain amplitude \cite{wyart08,tighe08a,sheinman12}, and steady flow \cite{tighe10c}.

\begin{acknowledgments}
This work was supported by the Dutch Organization for Scientific Research (NWO).
\end{acknowledgments}


\begin{thebibliography}{23}
\bibitem{footnote}{
We expect that increasing $\lambda_s$ increases $U^{\rm na}$ and that increasing $\lambda_f$ decreases $f$, both of which are consistent with resulting predictions that are verified numerically.}
\end{thebibliography}

\vspace{1cm}

{\bf Supplementary Material.---} 
For low frequency motions inertia can be neglected. The overdamped equations of motion can then be written \cite{tighe11}
\begin{equation}
{\bf K} {\bf Q} + \imath \omega {\bf B} {\bf Q} =  \sigma \Omega \hat \Gamma \,.
\end{equation} 
Here $\sigma$ is the shear stress, $\Omega$ is the volume, and $\hat \Gamma$ is a unit vector along the strain coordinate. ${\bf Q}$ is a vector comprising the $dN$ displacement components and the shear strain $\gamma$. ${\bf K}$ and ${\bf B}$ are stiffness and damping matrices, respectively, that quantify the elastic and viscous interactions. They are matrices of second derivatives
\begin{equation}
K_{mn}  = \frac{\partial^2 V}{\partial Q_m \partial Q_n} 
 \,\,\,\,\,\,\,\,{\rm and}\,\,\,\,\,\,\,\,
B_{mn} = \frac{\partial^2 R}{\partial \dot Q_m \partial \dot Q_n} \,,
 \end{equation}
 where $V$ is the elastic potential energy\begin{equation}
 V = \frac{1}{2} \sum_{\langle ij \rangle} k \, (u_{ij}^\parallel)^2 \,,
 \end{equation}
and $R$ is the Rayleigh dissipation function
\begin{equation}
R = \frac{1}{2} \sum_{i} b \, (\dot U^{\rm na}_i)^2
+ \frac{1}{2}b \dot \gamma^2 V\,.
\end{equation}

The numerical results reported in Figs.~3 and 4 are for periodic networks of $N = 256$ nodes. For each coordination number we average over approximately 100 networks.

The viscoelastic response can be characterized by studying the generalized eigenvalues and -vectors of $\lbrace {\bf K}, {\bf B}\rbrace$ \cite{tighe11}. The eigenvectors $\lbrace {\bf U}_n \rbrace$ are evanescent with relaxation rates given by the eigenvalues $\lbrace s_n \le 0 \rbrace$. Floppy modes have zero relaxation rate. Expanding $\bf Q$ in the eigenvectors and solving for the viscosity $\eta_0$ yields
\begin{equation}
\frac{N}{\eta_0} =  \sum_{(n | s_n = 0)} c_n +  \sum_{(n | s_n < 0)} \frac{c_n}{s_n} \,,
\end{equation}
The positive prefactor $c_n = |{\bf U}_n \cdot \hat \Gamma |^2/({\bf U}_n, {\bf B} {\bf U}_n)$ is the strength of the coupling between eigenmode $n$ and the imposed shear; it is positive and to very good approximation independent of $s_n$ \cite{tighe11,maloney06}. 
It is straightforward to show that the sum over floppy modes is the dominant contribution to the viscosity, so that
\begin{equation}
\eta_0 \sim \frac{N}{N_{\rm fm}} \sim \frac{1}{-\Delta z} \,,
\end{equation}
where $N_{\rm fm} = -\Delta z \, N/2$ is the number of floppy modes in the hypostatic system. Thus the number density of floppy modes directly sets the viscosity in hypostatic networks, similar to pressure fluctuations in jammed solids \cite{tighe10b}. The dynamic viscosity, or equivalently the scaling of non-affine fluctuations, can also be calculated in quasistatic hyperstatic networks, where loss is subdominant \cite{ellenbroek06,wyart08,tighe11}. One again finds $\eta_0 \sim 1/\Delta z$, consistent with the insets to Figs.~3a and b. Unlike the hypostatic case, however, these calculations require prior knowledge of either the static shear modulus or the form of the density of states.

\end{document}